# Soliton Blockade for Nonlinear Accelerating Pulses

Lifu Zhang,[1,*] Xuri Yang,[1] Qi Huang,[1] Yanxia Gao,[2] and Dianyuan Fan[1]

[1]International Collaborative Laboratory of 2D Materials for Optoelectronic Science & Technology, Institute of Microscale Optoelectronic Shenzhen University, Shenzhen 518060, China

[2]College of Physics and Optoelectronic Engineering, Shenzhen University, Shenzhen 518060, China

*Corresponding author: zhanglifu@szu.edu.cn

We study the nonlinear propagation of truncated Airyprime pulses in optical fibers with both anomalous or normal dispersion. We observe nonlinear self-accelerating pulses with notable red-shifted spectral notch (double peaks) or blue-shifted spectral peak depending on whether the dispersion is anomalous or normal. Such process is in sharp contrast to that of Airy pulses. The formation of nonlinear self-accelerating pulses is very sensitive to the truncated coefficient. The relationship between the characteristics of such accelerated pulses and the truncated coefficient are disclosed and compared in detail. Our results not only shed new light on the nonlinear propagation of Airyprime pulses, but also provide a novel method to generate nonlinear self-accelerating pulses as well as enable the realization of very efficient wavelength conversion based on the controlled frequency shift. Based on space-time duality, self-accelerating spatiotemporal nonlinear light bullets can be envisaged from the propagation of spatiotemporal Airyprime wave packets in pure Kerr medium.

Self-accelerating and non-spreading wave packets have attracted considerable attention since the optical Airy beams were first demonstrated experimentally and theoretically [1,2]. It is originated from the pioneering work by Berry and Balazs within the framework of quantum mechanics [3]. In contrast to the best known non-diffracting Bessel beam [4,5], the Airy beams not only have non-diffracting and self-healing but also lateral self-accelerating [1,2,6]. Such unique features stimulated extensive investigation of optical Airy beams and lead to a variety of applications such as supercontinuum generation [7], curved light filament [8], light bullets [9–11], microscopy [12,13], optical routing [14], laser micromaching [15], and *etc*.

Airy beams typically survive in linear media. While the Kerr nonlinearity that induced beams distortions tend to make their structures broken down [16,17]. However, Giannini and Joseph found self-accelerating solutions with Airy-like profiles in nonlinear Kerr media as early as 1989 [18]. Such nonlinear self-accelerating pulse can be used to realize synchrotron resonant radiation [19]. According to space-time duality, self-accelerating self-trapping beams were proved in nonlinear media with different nonlinear response such as saturable Kerr-like media, quadratic nonlinearities and nonlocal nonlinear media [20–23]. Interestingly, in the presence of weak nonlinearity, chirped Airy beams can evolves into nonlinear self-accelerating modes [24,25]. However, how to obtain a stable self-accelerating wave packets in strong nonlinearities is still a challenging work.

It is well known that Airy beams are described by Airy function [1,2]. Recently, Zhou et al. proposed a kind of beams expressed by Airyprime function, which is the first-order derivative of Airy function [26,27]. Compared with the Airy beams, the peak intensity of Airyprime beams is larger than that of Airy beams in free space propagation [28,29]. Consequently, the ring Airyprime beams have a powerful focusing ability compared with the ring Airy beams [30-33]. To date, related studies only focus on the linear regime. This raises many new intriguing questions. How the nonlinearity affects the Airyprime beams propagation? This question becomes even more interesting for temporal version of Airyprime beams (Airyprime pulses) propagation in optical fibers, where the group velocity dispersion is either normal or anomalous. Moreover, what happens to the Airyprime pulses nonlinear propagation in both cases? In addition, what are the differences between the Airy and Airyprime pulses nonlinear propagation? At a first analysis, one cannot expect that this asymmetric Airyprime pulses may lead to fairly non-trivial propagation dynamics, but, to the best of our knowledge, nonlinear Airyprime pulses propagation has not been reported before.

In this Letter, we investigate the nonlinear dynamics of truncated Airyprime pulses in both normal and anomalous dispersion regimes. Surprisingly, Airyprime pulses do not shed solitons but involve into self-accelerating pulses with red-shifted spectral notch or blue-shifted spectral peak, which depends on the sign of dispersion. This process is radically different from that of Airy pulses, in which shedding solitons appear.

The pulse propagation in an optical fiber can be described by the standard nonlinear Schrödinger equation (NLSE) [34]

$$i\frac{\partial U}{\partial \xi} = \frac{\operatorname{sgn}(\beta_2)}{2}\frac{\partial^2 U}{\partial \tau^2} - N^2|U|^2 U, \tag{1}$$

where $U(\tau,\xi)$ is normalized input. $\xi = z/L_D$ is normalized distance with dispersion length $L_D = |\beta_2|z/T_0^2$. $\tau = (t - z/v_g)/T_0$ is normalized with pulse duration $T_0$. $\beta_2$ represents the group velocity dispersion (GVD), and $v_g$ is the group velocity of the pulse. $\operatorname{sgn}(\Box)$ denotes the sign of GVD. $N = \sqrt{\gamma T_0^2/|\beta_2|}$ is soliton order. $\gamma$ is the nonlinear parameter.

The initial Airyprime pulsescan be written as follows:

$$U(\tau,\xi=0) = \sqrt{f(\alpha)} Ai'(\tau)\exp(\alpha\tau), \tag{2}$$

where $0 < \alpha < 1$ is the truncation coefficient, $f(\alpha)$ is the normalized coefficientto keep the pulse peak intensity to 1, and $Ai'(\tau)$ represents the first order derivate of Airy function. The initial spectrum of Airyprime pulses without normalized coefficient $f(\alpha)$ is:

$$U(\omega,0) = \sqrt{\alpha^2+\omega^2}\exp(\alpha^3/3)\exp(-\alpha\omega^2) \\ \times \exp\left[i\left(\alpha^2\omega - \omega^3/3 + \arctan(\omega/\alpha) + \pi\right)\right]. \tag{3}$$

Thus, we can obtain the initial spectral intensity and its derivative as:

$$I = |U(\omega,0)|^2 = (\alpha^2+\omega^2)\exp(2\alpha^3/3 - 2\alpha\omega^2). \tag{4}$$

According to Eq. (4), the initial spectrum is divided into two regions symmetrical about central frequency $\omega = 0$. The central frequency corresponds to an extreme point $I_{min} \approx \alpha^2$. The high-frequency and the low-frequency regions have same peak intensity $I_{max} \approx e^{-1}/2\alpha$ at $\omega_m \approx 1/\sqrt{2\alpha}$. Therefore, the smaller $\alpha$ the higher peak intensity and the greater spectral shift away from the center.

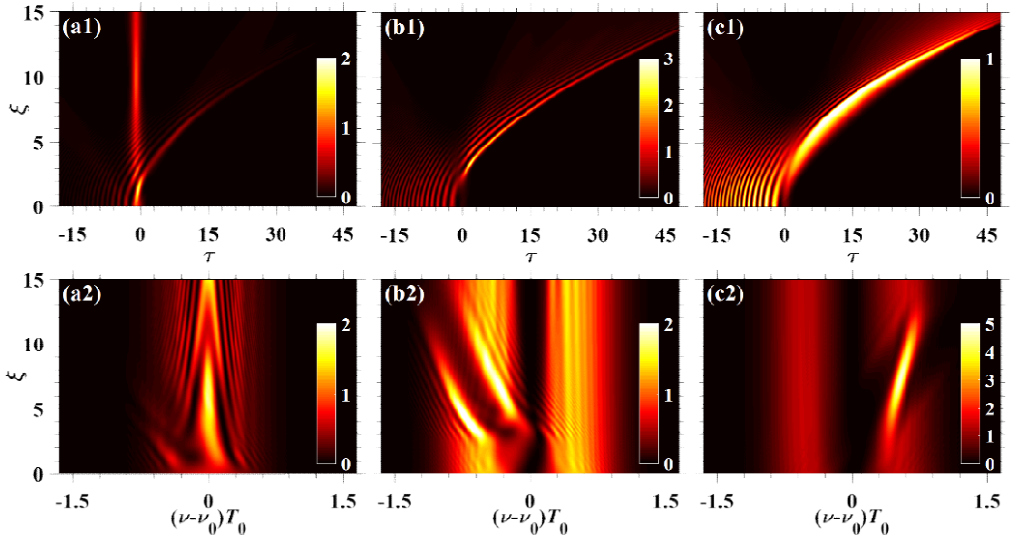

Fig. 1. Temporal and spectral evolution of Airy (a) and Airyprime (b, c) pulses in Kerr nonlinear media for anomalous (a, b) and normal (c) dispersion regimes.

Figures 1(b) and 1(c) show the temporal and spectral evolutions of the Airyprime pulses nonlinear propagation in anomalous and normal dispersion regimes, respectively. For a clear comparison, the Airy pulses nonlinear propagation in anomalous dispersion is plotted in Fig. 1(a). The parameters used in numerical simulations are $\alpha = 0.05$ and $N = 1.2$. When the Kerr nonlinearity is taken into account for the anomalous dispersion regime, it can be clearly seen from Figs. 1(a1) and 1(a2) that the Airy pulse is able to shed a soliton, and the residue part continues to accelerate; the corresponding spectrum experiences an obvious compression induced by self-phase modulation [16,17]. Airy pulses are quickly spreading in the normal dispersion regime (data not shown). While Figs. 1(b) and 1(c) show the Airyprime pulses exhibit completely different behaviors. In both cases, the Airyprime pulses keep their original structure during the initial stages of pulses evolution and then evolve into self-accelerating pulses. This transition is apparently seen from Figs. 1(b1) and 1(c1). The first lobe of self-accelerating pulse in the normal dispersion regime is much wider than that in anomalous dispersion regime. The corresponding spectral evolutions of Airyprime pulses shown in Figs. 1(b2) and 1(c2) indicate interesting features. For the case of anomalous dispersion, there is a red-shifted spectral notch located in the low frequency region during the process of self-accelerating pulse formation, while the high frequency region is almost unchanged. There is fundamental difference in the normal dispersion regime. An intensive spectral peak within high frequency region experiences blue shift during self-accelerating pulse propagation. Outside of the region of spectral gap or peak, the nonlinearity giver rises to a weak effect on the spectrum. Such controlled frequency shifts provide an alternative method to shaping spectrum for the application of wavelength conversion.

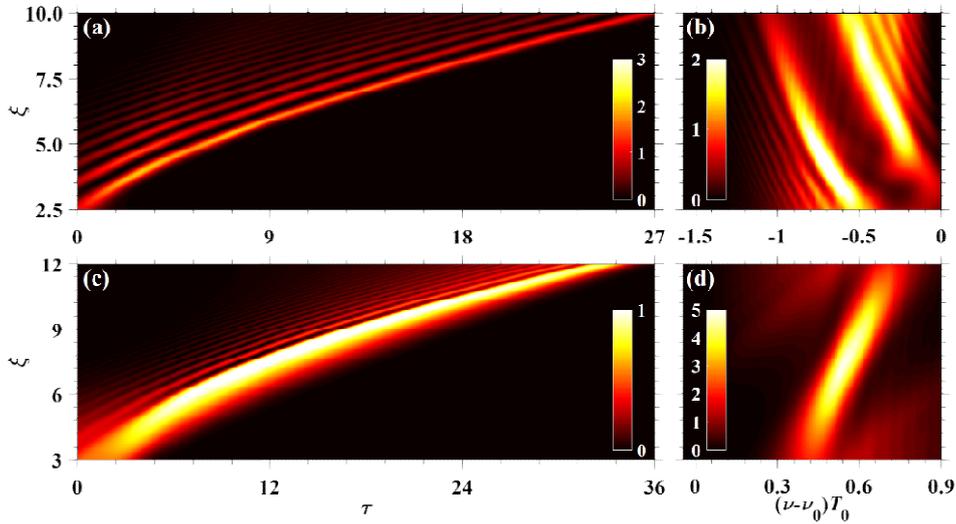

Fig. 2. Temporal and spectral evolution of the part of accelerated pulses shown in Figs. 1(b) and 1(c).

To further disclose the self-accelerating pulses formation, Figs. 2(a) and 2(c) show only the temporal and spectral evolution of the self-accelerated pulse for a zoomed vision of spectral notch or peak in anomalous and normal dispersion regimes respectively. It is reasonable to regard the pulses within this interval as ideal self-accelerating pulses. This is realized by filtering unimportant frequency components. The acceleration of the temporal trajectories and frequency shifts are evidently displayed. The differences between anomalous and normal dispersion regimes can be understood as follow. GVD causes the different frequency components to travel at slightly different speeds. Blue components travel faster than red components in the anomalous dispersion regime, while the opposite occurs in the normal dispersion regime. Moreover, accelerated pulses occupied low frequency components for keeping self-acceleration in the anomalous regime, while the opposite appears for the normal dispersion regime. The spectral shapes of Airyprime pulses were divided into low and high frequencies regions with respect to $\omega=0$. This is the key to form self-accelerating solutions. In addition, the peak intensity of the self-accelerated pulse formed in the anomalous dispersion is higher than that in normal dispersion, while the spectral peak intensity is opposite.

Like Airy pulses [1,2], the truncated coefficient is also the key parameter. It plays a relevant role and radically affects their propagation dynamics. The left and middle columns of Fig. 3 plotted the temporal and spectral evolution of Airyprime pulses with different truncated coefficient for both anomalous and normal dispersion regimes. For a comparison with Figs. 1(b) and 1(c) [$\alpha=0.05$], the truncated coefficient is chosen as $\alpha=0.03$ and $\alpha=0.07$. When $\alpha$ is decreased, the accelerated pulses are profoundly enhanced as shown in Figs. 3(a1) and 3(c1). While $\alpha$ is increased, Figs. 3(b1) and 3(d1) show that the characteristics of the accelerated pulses are significantly weakened. Consequently, the elongations of the spectral notch and peak are improved by decreasing the truncated coefficient. A series of numerical experiments were carried out for disclosing the relationship between the behaviors of the accelerated pulses and the truncated coefficient. The peak intensity ($I_M$) and its position ($\xi_p$), accelerating interval ($\Delta_\xi$), slope of frequency shift ($k_v$) are shown in the right column of Fig. 3. It can clearly see that these characteristic parameters are very sensitive to the truncated coefficient. The absolute value of the blue-shifted slope is much larger than that of red-shifted one. For the same truncated coefficient, the maximum intensity of spectral peak is larger than that of spectral notch, as shown in Fig. 3(e). When the truncated coefficient is increased to a critical value, Fig. 3(f) shows the distances required for reaching maximum are almost unchanged. The accelerating distance $\Delta_\xi$ decreases exponentially with an increasing of $\alpha$ [Fig. 3(g)]. For normal dispersion, the accelerating structure cannot be maintained as $\alpha>0.06$, while it is more robust for anomalous dispersion. For perfect nonlinear self-accelerating Airy pulses, their frequency shifts are linear [18,24,25]. Figure 3(h) plotted the slope of the frequency shifts $k_v$ as a function of truncated coefficient. The red shift is much faster than the speed of the blue shift. Their slopes are approximately linear, indicating the formation of nonlinear self-accelerating Airy pulses.

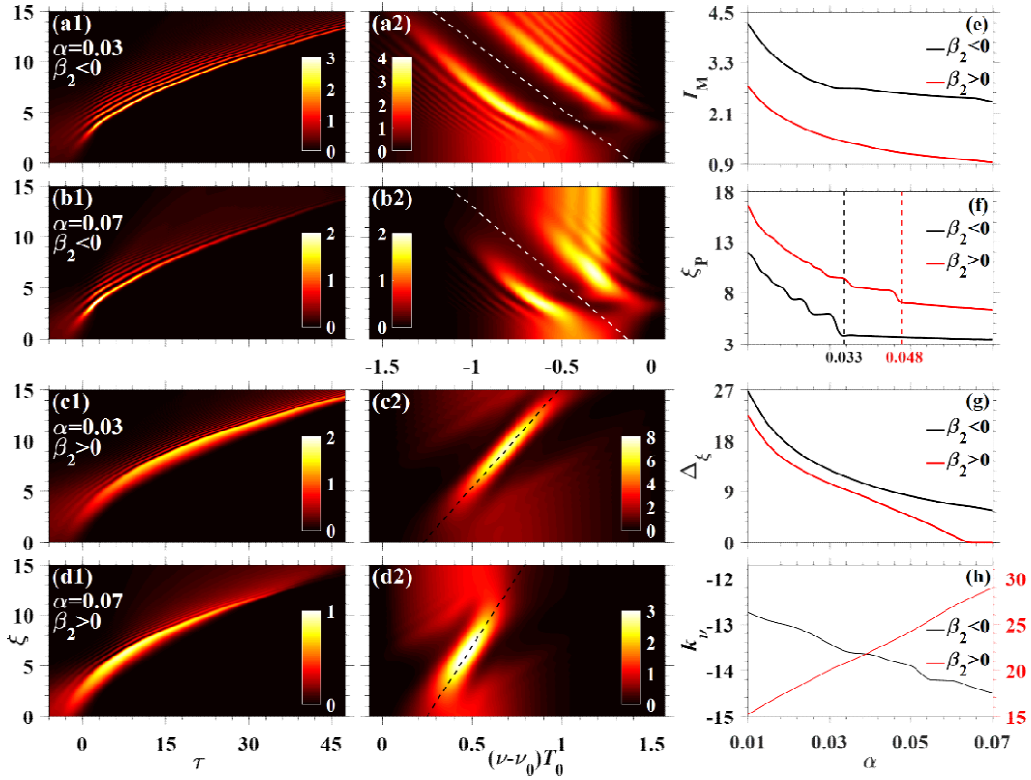

**Fig. 3.** Temporal (left column) and spectral (right column) evolutions of Airyprime pulses with different $\alpha$. Peak intensity (e), position of peak intensity (f), $\Delta_\xi$ (g) and $k_v$ (h) as a function of $\alpha$.

It is also necessary to investigate the stability under various degrees of nonlinearity. With an increasing nonlinearity, Airyprime pulse tends to shed energy from the intense lobes for keeping the shape of the accelerated pulses. As shown in Fig. 4(a), the dashed line depicts the trajectory of the peak intensity. The energy of the main lobe was transferred into the secondary lobes, which become obvious. The spectrum experiences a little bit distortion owing to the stronger nonlinearity [Fig. 4(b)]. Only the case for anomalous dispersion is illustrated because the stronger nonlinearity leads to serious dissipation during Airyprime pulse propagation n the normal dispersion regime. As illustrated in Figs. 4(c)-4(e), as $N$ is increased, the self-accelerating pulses formed in the normal dispersion regime were quickly broken down. As $N>1.4$, the accelerated distance $\Delta_\xi$ reaches zero. As a result, the propagation distance $\xi_P$ constantly shifts to an earlier distance with a transition [Fig. 4(d)]. While for the case of anomalous dispersion, the peak intensity first increases and reaches a maximum value, and then decreases when N is increased [Fig. 4(c)]. Even though the accelerating interval $\Delta_\xi$ is kept within a certain range of $\Delta_\xi=8$ [Fig. 4(e)], the propagation distance for approaching the peak intensity always decreases. The maximum values of $N$ without shedding energy is depicted in Fig. 4(f) versus the coefficient $\alpha$. $N_{max}$ decreases exponentially with an increasing $\alpha$, but not less than $N=1.4$. It is indicated that Airyprime pulses are more stable than Airy pulses in nonlinear propagation.

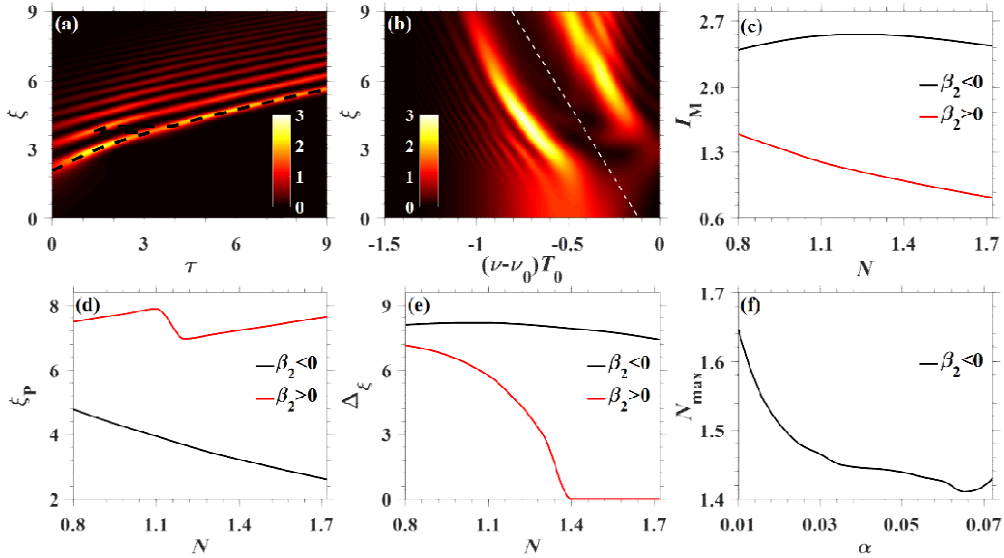

**Fig. 4.** (a) Temporal and (b) spectral evolution of Airyprime pulse for $N=1.5$. Peak intensity (c), position of peak intensity (d) and $\Delta_\xi$ (e) as a function of $N$. (h) $k_v$ as a function of $\alpha$.

In summary, we have investigated the nonlinear propagation of the Airyprime pulses in optical fibers with normal or anomalous dispersions. Compared with Airy pulses, the Airyprime pulses can evolve into self-accelerating pulses which are very sensitive to the truncation coefficient of Airyprime pulses. The spectrum of accelerated pulse exhibits only one blue-shifted intensive peak or double red-shifted peaks in normal or anomalous dispersion regimes. The differences of the self-accelerating pulses formed in both normal and anomalous dispersion regimes have been revealed and compared in detail. The robust of the accelerated pulses in anomalous dispersion is much higher than that in normal dispersion. Our findings not only deepen the understanding of Airyprime pulses nonlinear propagation, but also demonstrated a novel and simple method to generate nonlinear self-accelerating pulses. According to the space time duality, we believe to have put forth an excited route to generate self-accelerating spatiotemporal light bullets in nonlinear regime [35] beyond the linear light bullets [9,10,36-40].In addition, the nonlinear interaction of Airyprime pulses may be potential for photonic neuromorphic computing [41], rogue [42] and chaotic [43]optical wave dynamics.

This work was supported by the National NaturalScience Foundation of China (61975130), Guangdong Basic and Applied Basic Research Foundation (2021A1515010084).